\documentclass[aps,prl,superscriptaddress]{revtex4}
\usepackage{amsmath}
\usepackage{graphicx}
\usepackage{MnSymbol}
\usepackage{hyperref}
\usepackage{fancyhdr}
\usepackage{xcolor}

\begin{document}

\title{Interfacial fluid flow for systems with anisotropic roughness}

\author{B.N.J. Persson}
\affiliation{PGI-1, FZ J\"ulich, Germany, EU}
\affiliation{MultiscaleConsulting, Wolfshovener str 2, 52428 J\"ulich, Germany}

\begin{abstract}
I discuss fluid flow at the interface between solids with anisotropic roughness.
I show that for randomly rough surfaces with anisotropic roughness, the contact area
percolate at the same relative contact area as for isotropic roughness, and that
the Bruggeman effective medium theory and the critical junction theory give nearly the
same results for the fluid flow conductivity.
This shows that, in most cases, the surface roughness observed at
high magnification is irrelevant for fluid flow problems such as  
the leakage of static seals, and fluid squeeze-out.
\end{abstract}

\maketitle

\thispagestyle{fancy}

{\bf 1. Introduction}

Fluid flow at the interface between elastic solids is a complicated
topic, in general involving elastic deformations, 
complex fluid rheology and interfacial fluid slip\cite{slip}. 
In particular, the influence of the surface roughness on the fluid flow dynamics is a highly
complex topic. However, if there is a separation of length scales
the problem can be simplified: if
$R$ denote the (smallest) macroscopic radius of curvature of the (undeformed) 
surfaces in the nominal contact region, e.g., 
the radius of a ball, and
if $R>> \lambda_0$, where $\lambda_0$ is the longest (relevant) surface roughness component,
then it is possible to eliminate (integrate out) the surface roughness and obtain effective
fluid flow equations involving solid bodies with smooth surfaces (no roughness). 
The effective fluid flow equations depend on quantities determined by the surface roughness,
usually denoted fluid flow and friction factors (there are two fluid flow factors and three
friction factors). These factors depend on the average surface separation $\bar u$, which will vary 
throughout the nominal contact region; $\bar u$ is the local interfacial surface separation $u(x,y)$
averaged over the surface roughness\cite{YangP,Mart}. In several publications it has been shown how to calculate the
fluid flow factors, which enter in the (modified) Reynolds equation, and the friction factors, which 
enters in the expression for the shear stress acting on the solids\cite{Patir,Patir1,Patir2,slip,mic1,Salin,Tripp,Alm1}.

\begin{figure}
\includegraphics[width=0.35\textwidth,angle=0]{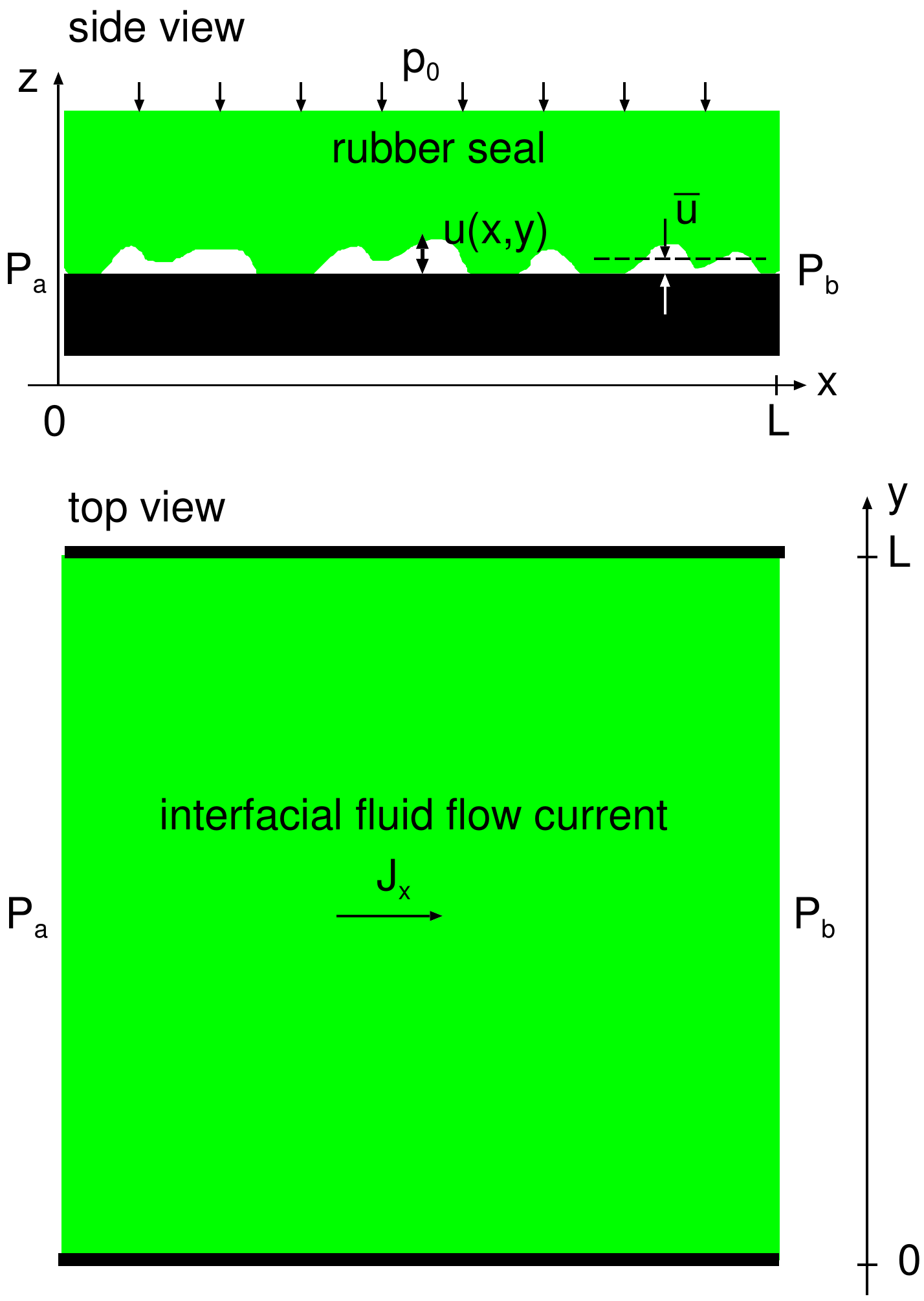}
\caption{\label{sealpic.pdf}
A square block $L\times L$ rubber seal (green) with surface roughness squeezed with the uniform pressure $p_0$
against a flat rigid countersurface. A fluid pressure difference $\Delta P = P_a-P_b >0$ occur between the two sides
$x=0$ and $x=L$. We assume no fluid leakage in the direction orthogonal to the $x$-axis i.e. two ends 
$y=0$ and $y=L$ are blocked.
}
\end{figure}

Here we consider the simplest fluid flow problems,
which include the leakage of static seals\cite{PY,Lorenz} and the squeeze-out of fluids\cite{squeeze} between
elastic solids. For these applications the roughness enter only via one function, namely the pressure flow factor
$\phi_{\rm p} (\bar u)$ (in general a $2\times 2$ tensor) or, equivalently, the (effective) fluid flow conductivity
$\sigma_{\rm eff}$ defined by the equation
$$\bar {\bf J} = -\sigma_{\rm eff} \nabla \bar p$$
where $\bar p=\langle p(x,y)\rangle$ is the fluid pressure and $\bar {\bf J}=\langle {\bf J}(x,y)\rangle$ 
the two-dimensional (2D) fluid flow current, 
both averaged over the surface roughness (ensemble averaging).
The flow conductivity $\sigma_{\rm eff}$ is a $2\times 2$ matrix (tensor). 

As an example, consider a seal consisting of a rubber block with square 
cross section $L\times L$, with surface roughness 
on length scales much smaller than $L$,
squeezed against a flat surface (see Fig. \ref{sealpic.pdf}).
Assume that high pressure fluid occur for $x<0$ and low pressure fluid for $x>L$ 
(pressure difference $\Delta P=P_{\rm a}-P_{\rm b} >0$).
In this case for the choosed coordinate system $\sigma_{\rm eff}$ is a diagonal matrix:
$$\sigma_{\rm eff} = \begin{pmatrix} \sigma_{x} & 0 \\0 & \sigma_{y}\end{pmatrix}$$
with $\sigma_y=0$.
The pressure gradient $\nabla \bar p$ is along the $x$-axis and
$d\bar p/dx = -\Delta P/L$. Thus, the fluid leakage rate (volume per unit time) becomes
$$\dot Q = L J_x =  L \sigma_x \Delta P/L = \sigma_x \Delta P$$

From the fluid flow conductivity one can calculate the pressure flow factor
$$\phi_{\rm p} = 12 \eta  \bar u^{-3}\sigma_{\rm eff}$$

For two parallel surfaces without roughness one has the flow conductivity (Poiseuille flow):
$$\sigma_0 = {u_0^3 \over 12 \eta},$$
where $u_0$ is the surface separation.
For a system with surface roughness it is sometimes convenient to define a separation $u_{\rm c}$,
which depends on the average surface separation $\bar u$, so that
$$\sigma_{\rm eff} = {u_{\rm c}^3 \over 12 \eta}$$
Thus the pressure flow factor
$$\phi_{\rm p} = \left ({u_{\rm c} \over \bar u} \right )^{3}$$

I this paper I discuss fluid flow at the interface between solids with anisotropic roughness.
I show that for randomly rough surfaces with anisotropic roughness, the contact area
percolate at the same relative contact area as for isotropic roughness. 
I also show that, unless the applied pressure is very small,
the Bruggeman effective medium theory and the critical junction theory give nearly the
same results for the fluid flow conductivity (and the fluid pressure flow factor).
This shows that for applications involves only the flow conductivity
(or, equivalently, the pressure flow factor), such as the leakage of static seals and 
fluid squeeze-out, in most cases the (short wavelength) 
surface roughness observed at high magnification is irrelevant.

\begin{figure}
\includegraphics[width=0.25\textwidth,angle=0]{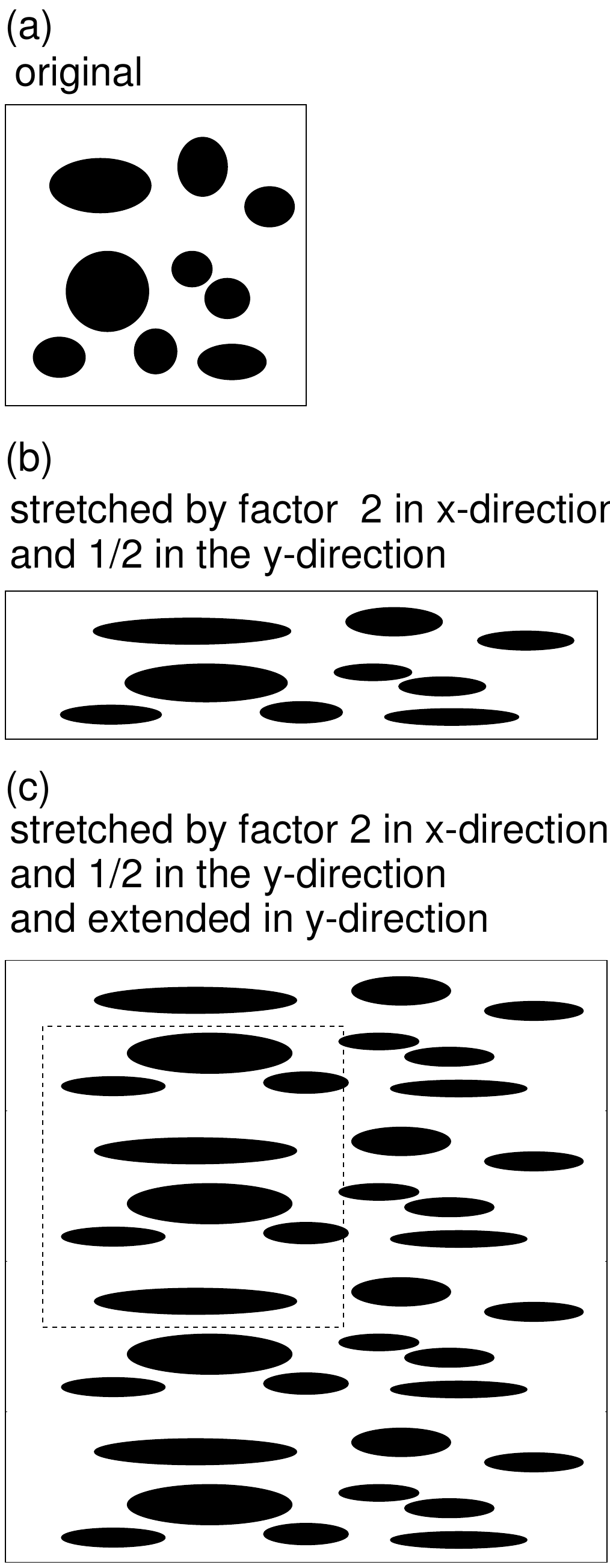}
\caption{\label{percolation.pdf}
(a) Asperity contact regions (black) for a system with random roughness with isotropic statistical properties, 
and (b), (c) for a system obtained by stretching by a factor of $2$ in the $x$-direction and 
$1/2$ in the y-direction. This transformation  conserve the area and result in anisotropic roughness with $\gamma = 4$.
For a system of finite size, if the system size is fixed [compare (a) with the dashed square in (c)] 
the contact area percolation threshold depends on the stretching factor $\gamma$, but
for an infinite system the percolation threshold does not depend on $\gamma$. 
}
\end{figure}

\vskip 0.3cm
{\bf 2. Qualitative discussion}

The theory of fluid flow discussed in this paper is based on the Bruggeman effective medium theory.
This theory is for an infinite-sized system but real applications and computer simulations 
involves systems of finite sizes. Finite size effects may in some cases be important, in particular for system
with strongly anisotropic roughness such as surfaces grinded in one direction, 
and for systems with small nominal contact area.

Consider the contact between two elastic solids with random surface roughness.
One way to (mathematically) produce systems with anisotropic roughness is to start with a 
surface with isotropic roughness, say a square area of size $L\times L$, and stretch 
the surface in the $x$-direction be a factor $\gamma^{1/2}$ and contract it
in the $y$-direction by a factor $\gamma^{-1/2}$, as
indicated in Fig. \ref{percolation.pdf}(b). 
This will map a circle on an ellipse (with the same surface area) where the ratio between the
ellipse axis in the $x$- and $y$-directions is gives by $\gamma$ 
(Peklenik number\cite{Peklenik}). 

To get a square unit surface we extend the surface in the $y$-direction
with similar rectangular units (but other realizations) as 
in Fig. \ref{percolation.pdf}(b), see Fig. \ref{percolation.pdf}(c).
If a surface region, with the same size $L\times L$ as the original surface, 
is cut-out of the surface in Fig. \ref{percolation.pdf}(c), 
the contact area may percolate in the $x$-direction 
(see dashed square in Fig. \ref{percolation.pdf}(c)) even if the contact area did not percolate for the original
surface. 
{\it However, for the infinite system the stretching cannot change the percolation threshold}. 
This is clear since a flow channel which is closed before stretching remain closed after stretching, and a flow
channel which is open before stretching will remain open after stretching. For 
a system with isotropic roughness the Bruggeman theory predict that the contact area percolate when
$A/A_0 = 0.5$, and the same is true for a surface with anisotropic roughness (see Sec.5). 

In Ref. \cite{Ref2} I presented an approximate formula for the fluid flow conductivity which interpolate between the
Bruggeman effective medium theory result for isotropic roughness, and the known limit for the fluid flow conductivity
for the case of strongly anisotropic roughness. 
The expression for the flow conductivity $\sigma_x$
proposed in Ref. \cite{Ref2} is
$${1\over \sigma_x} = \left \langle {1+\gamma \over \sigma+\gamma \sigma_x}\right \rangle\eqno(1)$$
where $\sigma = u^3/(12 \eta)$, where $\eta$ is the fluid viscosity and $u=u({\bf x})$ the interfacial separation 
at the point ${\bf x}=(x,y)$.  The $\langle .. \rangle$ stands for ensemble averaging, or averaging over the
probability distribution $P(u)$ of interfacial separations.
For $\gamma = 1$ this equation reduces to the standard Bruggeman equation for isotropic roughness,
while for $\gamma \rightarrow 0$ it gives $\sigma_x = \langle \sigma^{-1} \rangle^{-1}$ and for 
$\gamma \rightarrow \infty$ it gives $\sigma_x = \langle \sigma\rangle$. Both these limits are exact
results as is easy to show directly from the Reynolds equation for thin film fluid flow. Nevertheless,
for an infinite system (1) gives the wrong percolation condition (see below). 

The flow conductivity in the $y$-direction is obtained from (1) by replacing $\gamma$ with $1/\gamma$:
$${1\over \sigma_y} = \left \langle {1+(1/\gamma) \over \sigma+(1/\gamma) \sigma_y}\right \rangle\eqno(2)$$

We can write the probability distribution of interfacial separation as\cite{Alm,Carb1}
$$P(u)= {A\over A_0} \delta (u) + P_{\rm c}(u)\eqno(3)$$
where $P_{\rm c}(u)$ is the (continuous) part of the distribution where $u>0$. Thus we get
$${1\over \sigma_x} = {1+\gamma \over \sigma_x \gamma} {A\over A_0}+  \left \langle {1+\gamma\over \sigma+\gamma \sigma_x}
\right \rangle_{\rm c}\eqno(4)$$
When the contact area percolate no fluid 
flow is possible from one side to the other side of the studied unit, so that
$\sigma_x \rightarrow 0$. 
When $\sigma_x \rightarrow 0$ using (4) gives
$$1={1+\gamma \over \gamma} {A\over A_0}$$
or $A/A_0 = \gamma /(1+\gamma)$. However, for an infinite system this result is incorrect and the relative contact area $A/A_0$
at the point where the contact area percolate does in fact not depend on $\gamma$. 

Computer simulations of contact mechanics are always for finite-sized systems. In this case it has been observed
that when $\gamma > 1$ the contact area percolate for a smaller relative contact area $A/A_0$ then when
$\gamma = 1$. Similarly, for $\gamma < 1$ the contact area percolate for a larger $A/A_0$ then when
$\gamma = 1$, and in one study the results was rather accurately described by the formula $A/A_0 = \gamma /(1+\gamma)$. 
These results are intuitively clear, but the simulation results depend on the system size, and is hence non-universal. 

The Bruggeman effective medium theory gives flow 
conductivities of the form (1) and (2), but with $\gamma$ replaced by $\gamma^* = \gamma (\sigma_y/\sigma_x)^{1/2}$
(see Sec. 4). For this case the percolation threshold occur when $A/A_0 = 0.5$ independent of $\gamma$ (see Sec. 5).

\vskip 0.3cm
{\bf 3 Tripp number}

The most important property characterizing a rough surface is the surface roughness power spectrum $C({\bf q})$.
If $z=h({\bf x})$ is the height coordinate at the point ${\bf x} = (x,y)$
then the two-dimensional (2D) power spectrum $C({\bf q}) = C(q_x,q_y)$ is given by 
$$C({\bf q}) = {1\over (2 \pi )^2} \int d^2x \ \langle  h({\bf x})h({\bf 0}) \rangle e^{i {\bf q} \cdot {\bf x}}\eqno(5)$$
where $\langle .. \rangle$ stands for ensemble averaging.
For a surface with isotropic statistical properties, $C({\bf q})$ depends only on the magnitude $q=|{\bf q}|$ of the 2D wave vector
${\bf q}$. For surfaces with anisotropic statistical properties the Tripp number\cite{Tripp} $\gamma (q)$ is very important as it
determines the influence of the surface roughness anisotropy on
interfacial fluid flow\cite{Tripp,JPC}. The Tripp number depends on the length scale considered, i.e., it is a function of the wavenumber $q$,
and is defined as follows\cite{JPC}.
We introduce polar coordinates ${\bf q} = q ({\rm cos}\phi, {\rm sin}\phi)$ and define the matrix
$$D(q) = {\int_0^{2\pi} d\phi \ C({\bf q}) {\bf q} {\bf q}/q^2 \over \int_0^{2\pi} d\phi \ C({\bf q}) }\eqno(6)$$
Note that $D(q)$ is a symmetric matrix and can be diagonalized by an orthogonal transformation.
We denote the diagonal elements by $1/(1+\gamma)$ and $\gamma / (1+\gamma)$ where
$\gamma = \gamma(q)$ is the Tripp number, which depends on the  wavenumber $q$. If $C({\bf q})$ only depend on the
magnitude of the wavevector then $D_{ij}(q)=\delta_{ij}/2$, so that $\gamma = 1$ for
roughness with isotropic statistical properties.

One can also define the average Tripp number using 
$$D ={  \int d^2q \ C({\bf q}) {{\bf q} {\bf q}/ q^2}\over \int d^2q \ C({\bf q}) }\eqno(7)$$
Let us study 
$$I = \int d^2q \ C({\bf q}) {{\bf q} {\bf q} \over q^2}$$
for a particular case. Assume that $f(x,y) = f(r)$ only depend on the magnitude of the coordinate ${\bf x}$ and consider the function $f(x/a_x,y/a_y)$.
If $f(r)=0$ is a circle then $f(x/a_x,y/a_y)=0$ is an ellipse. In wavevector space we get the function $g(q_xa_x,q_ya_y)$.
Assume that $C({\bf q}) = g(q_x a_x,q_y a_y)$ with $g(q_x,q_y)=g(q)$. Writing
$q_x'=q_x a_x$, $q_y'=q_y a_y$ we get
$$I= {1\over a_x a_y} \int d^2q' \ g(q') {1\over (q_x'/a_x)^2+(q_y'/a_y)^2} 
  \left( {\begin{array}{cc}
   (q_x'/a_x)^2 & q_x'q_y'/(a_xa_y) \\
   q_x'q_y'/(a_xa_y) & (q_y'/a_y)^2 \\
  \end{array} } \right)
$$
In polar coordinates 
$$q_x'=q' {\rm cos} \phi, \ \ \ \ \ \ q_y'=q' {\rm sin} \phi$$
we get
$$I= {1\over a_x a_y} \int d^2q' \ g(q') {1\over a_x^{-2} {\rm cos}^2\phi +a_y^{-2} {\rm sin}^2\phi} 
  \left( {\begin{array}{cc}
   a_x^{-2} {\rm cos}^2\phi &  (a_xa_y)^{-1} {\rm cos}\phi \ {\rm sin} \phi \\
   (a_xa_y)^{-1} {\rm cos}\phi \ {\rm sin} \phi  & a_y^{-2} {\rm sin}^2\phi  \\
  \end{array} } \right)
$$
or
$$I= {1\over a_x a_y} \int d^2q' \ g(q') {1\over {\rm cos}^2\phi +\gamma^2 {\rm sin}^2\phi} 
  \left( {\begin{array}{cc}
   {\rm cos}^2\phi &  0 \\
   0  & \gamma^2 {\rm sin}^2\phi  \\
  \end{array} } \right) \eqno(8)$$
Let us denote $a_x/a_y = \gamma$ which we refer to as the Tripp number.
Since $g(q')$ only depend on the magnitude of ${\bf q}'$ we can write (8) as
$$I= {1\over a_x a_y} \int d^2q' \ g(q') {1\over 2 \pi} \int d\phi {1\over {\rm cos}^2\phi +\gamma^2 {\rm sin}^2\phi} 
  \left( {\begin{array}{cc}
   {\rm cos}^2\phi &  0 \\
   0  & \gamma^2 {\rm sin}^2\phi  \\
  \end{array} } \right)
$$
Finally using that 
$${1\over a_x a_y} \int d^2q' \ g(q') = \int d^2q \ g(q_x a_x, q_y a_y) =  \int d^2q \ C({\bf q})$$
we get 
$$D=  
{1\over 2 \pi} \int d\phi {1\over {\rm cos}^2\phi +\gamma^2 {\rm sin}^2\phi} 
  \left( {\begin{array}{cc}
   {\rm cos}^2\phi &  0 \\
   0  & \gamma^2 {\rm sin}^2\phi  \\
  \end{array} } \right)
$$
The integral over $\phi$ is easy to perform (see Appendix A) giving
$$D= {1\over 1+\gamma}   
  \left( {\begin{array}{cc}
   1 &  0 \\
   0  & \gamma  \\
  \end{array} } \right)
\eqno(9)$$

\begin{figure}
\includegraphics[width=0.45\textwidth,angle=0]{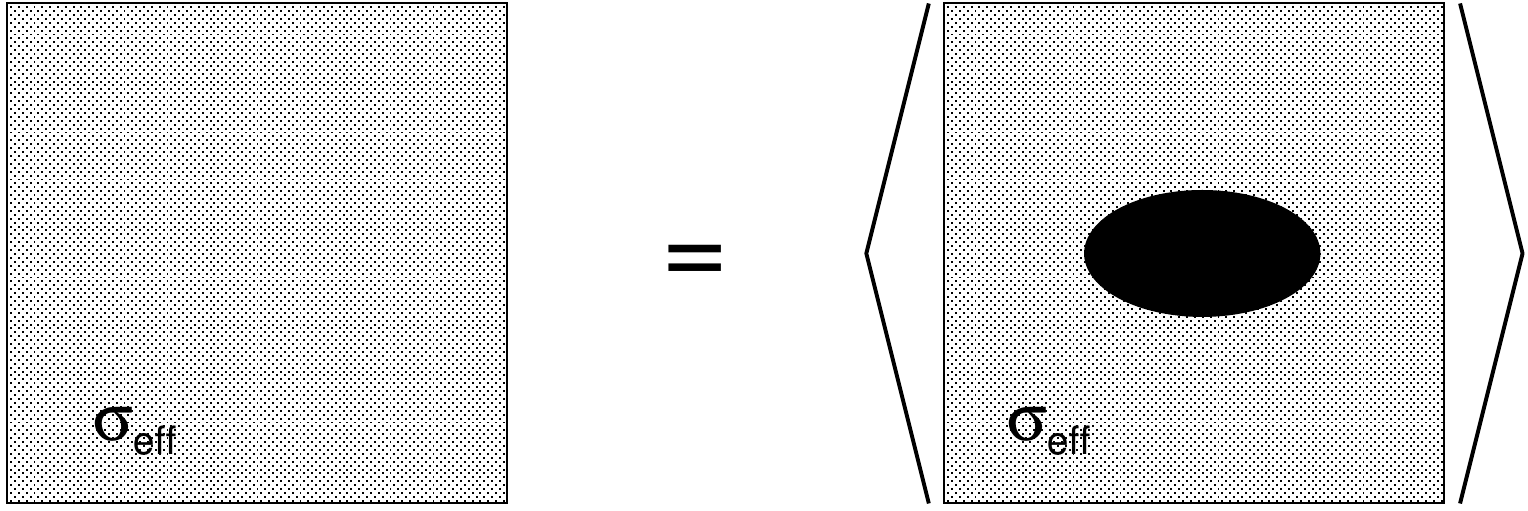}
\caption{\label{EffectiveMedium.pdf}
In the effective medium approach a system with anisotropic roughness
is replaced by an effective system with the (constant) flow conductivity $\sigma_{\rm eff}$.
The effective flow conductivity is determined as follows: An elliptic region with the constant
surface separation $u$ is embedded in the effective medium. The flow current ${\bf J}_u(x,y)$
for this system depends on the surface separation $u$ in the elliptic region.
The effective conductivity is determined by the condition that 
${\bf J}_u $ averaged over the probability 
distribution of surface separations $P(u)$ is equal to the 
flow current obtained using the effective medium everywhere.
}
\end{figure}

\vskip 0.3cm
{\bf 4 Bruggeman effective medium theory for fluid flow}

Effective medium theories are simple, but very useful and often accurate methods to describe 
some properties of inhomogeneous materials. The effective medium approach assumes
that the material in randomly disordered at length scales much shorter than the
the length scale of interest. Typical applications of effective medium theories are
the optical properties of inhomogeneous materials, and the electric or fluid transport in
inhomogeneous materials. There are several different (but related) 
effective medium theories, e.g. the coherent potential
approximation or the Bruggeman effective medium approximation\cite{Brug,Ref1,Scar1,Scar,A1,A2,A3,Dapp}. 
In earlier publications we have shown how the leakage of seals can be accurately 
described using the Bruggeman effective medium theory for systems with random but 
isotropic surface roughness\cite{Lorenz}. 

The fluid flow current
$${\bf J} = - \sigma \nabla p\eqno(10)$$
where 
$$\sigma = {u^3 \over 12 \eta}\eqno(11)$$
where $u({\bf x})$ is the interfacial separation at the point ${\bf x}$ and $\eta$ the fluid viscosity.
Conservation of mass 
$$\nabla \cdot {\bf J} = 0\eqno(12)$$
We will replace the inhomogeneous system with a homogeneous system with the average interfacial
separation $\bar u ({\bf x})$ which can be treated as locally a constant. The average flow current
$$\bar {\bf J}  = - \sigma_{\rm eff} \nabla \bar p\eqno(13)$$

The flow conductivity $\sigma_{\rm eff}$ in the Bruggeman effective medium 
approach is determined as indicated in Fig. \ref{EffectiveMedium.pdf}. 
That is, in the effective medium approach a system with anisotropic roughness
is replaced by an effective system with the (constant) flow conductivity $\sigma_{\rm eff}$.
The effective flow conductivity is determined as follows: An elliptic region with the constant
surface separation $u$ is embedded in the effective medium. The flow current ${\bf J}_u(x,y)$
for this system depends on the surface separation $u$ in the elliptic region.
The effective conductivity is determined by the condition that 
${\bf J}_u $ averaged over the probability 
distribution of surface separations $P(u)$ is equal to the 
flow current obtained using the effective medium everywhere.

The treatment which follows is similar to those presented in Ref. \cite{Ref1} and \cite{Scar1}.
Let us write
$${\bf J}_u = -\sigma \nabla p \ \ \ \ \ \ {\rm inside \ the \ elliptic \ region}\eqno(14)$$
$${\bf J}_u = -\sigma_{\rm eff} \nabla p \ \ \ \ \ \ {\rm outside \ the \ elliptic \ region}\eqno(15)$$
Thus if we define 
$${\bf J}_u = - \sigma_{\rm eff} \nabla p + {\bf J}_1\eqno(16)$$
then ${\bf J}_1 = {\bf 0}$ outside the elliptic region.
Using (12) we get
$$\nabla \cdot \sigma_{\rm eff} \nabla p  = \nabla \cdot {\bf J}_1 = 
\int d^2x' \delta ({\bf x}-{\bf x}') \nabla \cdot {\bf J}_1 ({\bf x}') \eqno(17)$$
If we define
$$\nabla \cdot \sigma_{\rm eff} \nabla G({\bf x}-{\bf x}') =  \delta ({\bf x}-{\bf x}')\eqno(18)$$
we can write
$$\nabla \cdot \left [ \sigma_{\rm eff} \left (\nabla p - \int d^2x' \nabla G({\bf x}-{\bf x}') 
\nabla' \cdot {\bf J}_1 ({\bf x}') \right ) \right ] = 0$$
This equation is satisfied by
$$\nabla p - \int d^2x' \nabla G({\bf x}-{\bf x}') \nabla' \cdot {\bf J}_1 ({\bf x}') =\nabla p^o$$
where $\nabla p^o$ is a constant vector. Thus
$$\nabla p =\nabla p^o+ \int d^2x' \nabla \nabla G({\bf x}-{\bf x}') \cdot {\bf J}_1 ({\bf x}')\eqno(19)$$
where we have performed a partial integration and used that ${\bf J}_1 ({\bf x})$ vanish outside the elliptic region.

The problem above involves equations very similar to those in electrostatics
(see Appendix B). From electrostatics we know that
if an elliptic (homogeneous) body is embedded in a (homogeneous) dielectric media, 
in an applied electric field the electric polarization 
in the inclusion is uniform. In the present case this imply that ${\bf J}_1$ is constant in the elliptic region where the interfacial
separation equals $u$ (a constant). Since ${\bf J}_1$ vanish outside the elliptic region, if 
we define $f({\bf x})=1$ inside the elliptic region and $f({\bf x})=0$ outside, we can write (19) ads 
$$\nabla p = \nabla p^o + Q \cdot {\bf J}_1 \eqno(20)$$ 
where the matrix
$$Q=\int d^2x \ f({\bf x}) \nabla \nabla G({\bf x})\eqno(21)$$
where the integral is over the whole $xy$-plane.
Since ${\bf J}_u = -\sigma \nabla p$ inside the elliptic region (see (14)) from (16) we get
$${\bf J}_1 = (\sigma_{\rm eff}-\sigma ) \nabla p\eqno(22)$$
Substituting this in (20) gives
$$\nabla p = \nabla p^o + Q  \cdot (\sigma_{\rm eff}-\sigma ) \nabla p$$
or
$$\left [1-Q  \cdot (\sigma_{\rm eff}-\sigma ) \right ] \nabla p = \nabla p^o$$
or
$$\nabla p =  \left [1-Q  \cdot (\sigma_{\rm eff}-\sigma ) \right ]^{-1}   \nabla p^o\eqno(23)$$
Using (22) and (23) we get
$${\bf J_1} = (\sigma_{\rm eff} -\sigma ) \nabla p =  (\sigma_{\rm eff} -\sigma ) 
\left [1-Q  \cdot (\sigma_{\rm eff}-\sigma ) \right ]^{-1}   \nabla p^o$$
We demand that the average of ${\bf J_1}$ vanish which gives
$$\langle {\bf J_1} \rangle = \langle (\sigma_{\rm eff}-\sigma ) \left [1-Q  \cdot (\sigma_{\rm eff}-\sigma ) \right ]^{-1} 
\rangle  \nabla p^o =0$$
Since $\nabla p^o$ is an arbitrary constant vector we get
$$\langle (\sigma_{\rm eff}-\sigma ) \left [1-Q  \cdot (\sigma_{\rm eff}-\sigma ) \right ]^{-1} \rangle  =0\eqno(24)$$
Since $Q$ is a diagonal matrix (see below) with components $Q_{11}$ and $Q_{22}$, the matrix
$M=[1-Q  \cdot (\sigma_{\rm eff}-\sigma )]$ is also diagonal with the elements
$$M_{11}= 1-Q_{11} (\sigma_x - \sigma)$$
and
$$M_{22}= 1-Q_{22} (\sigma_y - \sigma)$$
Thus we get from (24):
$$\left \langle {\sigma_x-\sigma \over 1-Q_{11} (\sigma_x - \sigma)} \right \rangle =0\eqno(25)$$
$$\left \langle {\sigma_y-\sigma \over 1-Q_{22} (\sigma_y - \sigma)} \right \rangle =0\eqno(26)$$

The Fourier transform of (18) gives
$$- {\bf q} \cdot \sigma_{\rm eff} {\bf q} G({\bf q}) =  {1\over (2 \pi )^2}$$
or
$$G({\bf q}) = - {1\over (2 \pi )^2} {1 \over \sigma_x q_x^2 +\sigma_y q_y^2}\eqno(27)$$
and (20) gives
$$Q=\int d^2x \int d^2q \ d^2q' \ f({\bf q'})  (- {\bf q} {\bf q}) G({\bf q}) e^{i({\bf q}+{\bf q'})\cdot {\bf x}}$$
$$= (2\pi )^2 \int d^2q \ f({\bf q})  (- {\bf q} {\bf q}) G({\bf q})\eqno(28)$$
Using (27) and  (28) and  $q_x' = q_x a_x$, $q_y' = q_y a_y$ and using that $f({\bf q}) = f(q')$ we get
$$Q={1 \over a_x a_y} \int d^2q' \ f(q') {1\over \sigma_x (q_x'/a_x)^2+\sigma_y (q_y'/a_y)^2} 
  \left( {\begin{array}{cc}
   (q_x'/a_x)^2 & q_x'q_y'/(a_xa_y) \\
   q_x'q_y'/(a_xa_y) & (q_y'/a_y)^2 \\
  \end{array} } \right)
$$
or
$$Q={1 \over a_x a_y} \int d^2q' \ f(q') {1\over 2 \pi} \int d\phi \ {1\over \sigma_x a_x^{-2} {\rm cos}^2 \phi+\sigma_y a_y^{-2} {\rm sin}^2 \phi} 
  \left( {\begin{array}{cc}
   a_x^{-2} {\rm cos}^2\phi &  (a_xa_y)^{-1} {\rm cos}\phi \ {\rm sin} \phi \\
   (a_xa_y)^{-1} {\rm cos}\phi \ {\rm sin} \phi  & a_y^{-2} {\rm sin}^2\phi  \\
  \end{array} } \right)
$$
Using that 
$${1 \over a_x a_y} \int d^2q' \ f(q') = \int d^2q \ f({\bf q}) = f({\bf x=0}) = 1$$
we get
$$Q={1\over 2\pi }  \int d\phi \ {1\over \sigma_x a_x^{-2} {\rm cos}^2 \phi+\sigma_y a_y^{-2} {\rm sin}^2 \phi} 
  \left( {\begin{array}{cc}
   a_x^{-2} {\rm cos}^2\phi &  (a_xa_y)^{-1} {\rm cos}\phi \ {\rm sin} \phi \\
   (a_xa_y)^{-1} {\rm cos}\phi \ {\rm sin} \phi  & a_y^{-2} {\rm sin}^2\phi  \\
  \end{array} } \right)
$$
$$={1 \over 2\pi \sigma_x} \int d\phi \ {1\over{\rm cos}^2 \phi+\gamma^2 (\sigma_y/\sigma_x)  {\rm sin}^2 \phi}
  \left( {\begin{array}{cc}
  {\rm cos}^2\phi &  0 \\
   0  & \gamma^2 {\rm sin}^2\phi  \\
  \end{array} } \right)
$$
Using (A1) and (A2) this gives
$$Q_{11}=
  {1\over \sigma_x}{1\over 1+\gamma (\sigma_y/\sigma_x)^{1/2}}\eqno(29) 
$$
$$Q_{22}= 
   {1\over \sigma_y}{\gamma (\sigma_y/\sigma_x)^{1/2} \over 1+\gamma (\sigma_y/\sigma_x)^{1/2}} \eqno(30)  
$$
Substituting (29) in (25) and (30) in (26) gives
$${1\over \sigma_x} = \left \langle {1+\gamma^*\over \sigma+\gamma^* \sigma_x}\right \rangle\eqno(31)$$
$${1\over \sigma_y} = \left \langle {1+(1/\gamma^*)\over \sigma+(1/\gamma^*) \sigma_y}\right \rangle\eqno(32)$$
where $\gamma^* = \gamma (\sigma_y/\sigma_x)^{1/2}$.

\begin{figure}
\includegraphics[width=0.25\textwidth,angle=0]{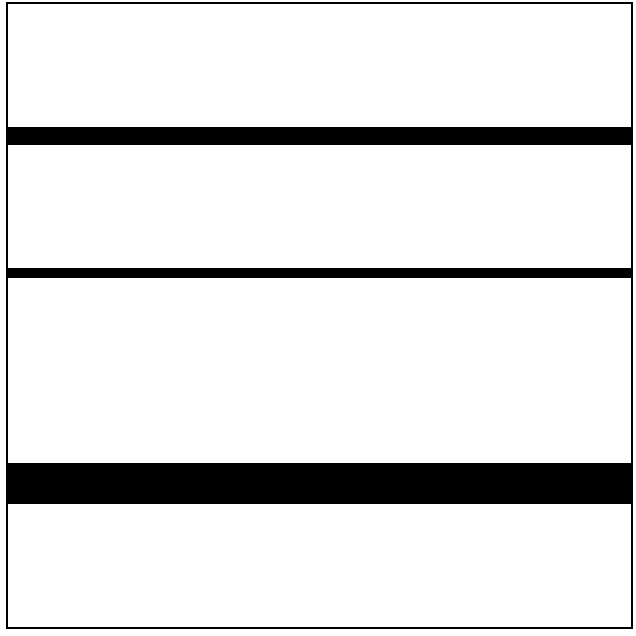}
\caption{\label{strip}
Example of area of real contact (black) as $\gamma \rightarrow \infty$.
}
\end{figure}

\vskip 0.3cm
{\bf 5 Limiting cases}

Consider first the case when $\gamma^* \rightarrow \infty$. In this case from (31) we get $\sigma_x = \langle \sigma \rangle$ and
from (32) $\sigma_y = \langle \sigma^{-1} \rangle^{-1}$. Note that 
$$\gamma^* = \gamma \left ({\sigma_y \over \sigma_x}\right )^{1/2} = 
{\gamma \over \left (\langle \sigma \rangle \langle \sigma^{-1} \rangle \right )^{1/2}}$$ 
Thus if $\gamma \rightarrow \infty$ it follows that $\gamma^* \rightarrow \infty$.

Note that when  $\gamma^* \rightarrow \infty$, if the area of real contact $A>0$ we get $\langle u^{-1} \rangle^{-1} = 0$, so 
that $\sigma_y=0$ and no fluid can flow in
the $y$-direction. This result is clear from a physical point of view since strips of 
contact will extend between the two edges of the system in the $x$-direction as indicated
in Fig. \ref{strip} and no fluid flow is possible. The results for $\sigma_x$ and $\sigma_y$ when 
$\gamma \rightarrow \infty$ (or $\gamma \rightarrow 0$) are well known and can be easily obtained directly from
the Reynold thin-film fluid flow equation with $u(x,y)$ only depending on $y$ (or $x$).

Next, let us consider the case when we increase the nominal contact pressure so we approach the limit when 
the contact area percolate. 
When the contact area percolate no fluid 
flow is possible from one side to the other side of the studied unit, so that
$\sigma_x \rightarrow 0$. 
We can write the probability distribution of interfacial separation as\cite{Alm,Carb1}
$$P(u)= {A\over A_0} \delta (u) + P_{\rm c}(u)$$
where $P_{\rm c}(u)$ is the part of the distribution where $u>0$. Thus we get
$${1\over \sigma_x} = {1+\gamma^* \over \sigma_x \gamma^*} {A\over A_0}+  \left \langle {1+\gamma^*\over \sigma+\gamma^* \sigma_x}
\right \rangle_{\rm c}\eqno(33)$$
$${1\over \sigma_y} =  {1+\gamma^*\over \sigma_y} {A\over A_0}+ \left \langle {1+(1/\gamma^*)\over 
\sigma+(1/\gamma^*) \sigma_y}\right \rangle_{\rm c}\eqno(34)$$
where $\langle .. \rangle_{\rm c}$ stands for averaging using $P_{\rm c}(u)$ i.e., over the non-contact surface area $A_0-A$.
From (33) as $\sigma_x \rightarrow 0$ we get 
$$1={1+\gamma^* \over  \gamma^*} {A\over A_0}\eqno(35)$$
We will now show that $\sigma_x \rightarrow 0$ imply $\sigma_y \rightarrow 0$ i.e. the contact area percolate in both the $x$ and $y$-directions
at the same time. To prove this, assume that this is not the case so $\sigma_y$ remains non-zero as
$\sigma_x \rightarrow 0$. It then follows that $\gamma^* = \gamma (\sigma_y/\sigma_x)^{1/2} \rightarrow \infty$ as $\sigma_x \rightarrow 0$.
In this case (35) gives $A/A_0 = 1$. This result is incorrect because we know that the contact area when $\gamma = 1$ percolate when
$A/A_0 = 0.5$. Thus $\sigma_x \rightarrow 0$ imply $\sigma_y \rightarrow 0$ and
from (34) we get
$$1=(1+\gamma^*) {A\over A_0}\eqno(36)$$
Using (35) and (36) gives
$\gamma^*=1$ and $A/A_0 = 1/2$. Using $\gamma^* = \gamma (\sigma_y/\sigma_x)^{1/2}$ and $\gamma^*=1$
we get $\sigma_x = \gamma^2 \sigma_y$, which holds as $\sigma_x \rightarrow 0$.

Finally, let us consider the case when the separation $u({\bf x})=\bar u +\delta u({\bf x})$ where $\bar u$ is the average separation
and $\delta u / \bar u << 1$. This case was studied in Appendix A in Ref. \cite{Ref2} but where we now must replace $\gamma$ with $\gamma^*$.
However, since $\gamma^* = \gamma$ to zero order in $\delta u$ the results derived in Appendix A in Ref. \cite{Ref2} are still valid
and we conclude that the effective medium theory result for $\sigma_x$ and $\sigma_y$ is exact to order $\delta u^2$ and that
$\gamma$ can be obtained from the matrix $D$ involving only the surface roughness power spectrum.

\begin{figure}
\includegraphics[width=0.35\textwidth,angle=0]{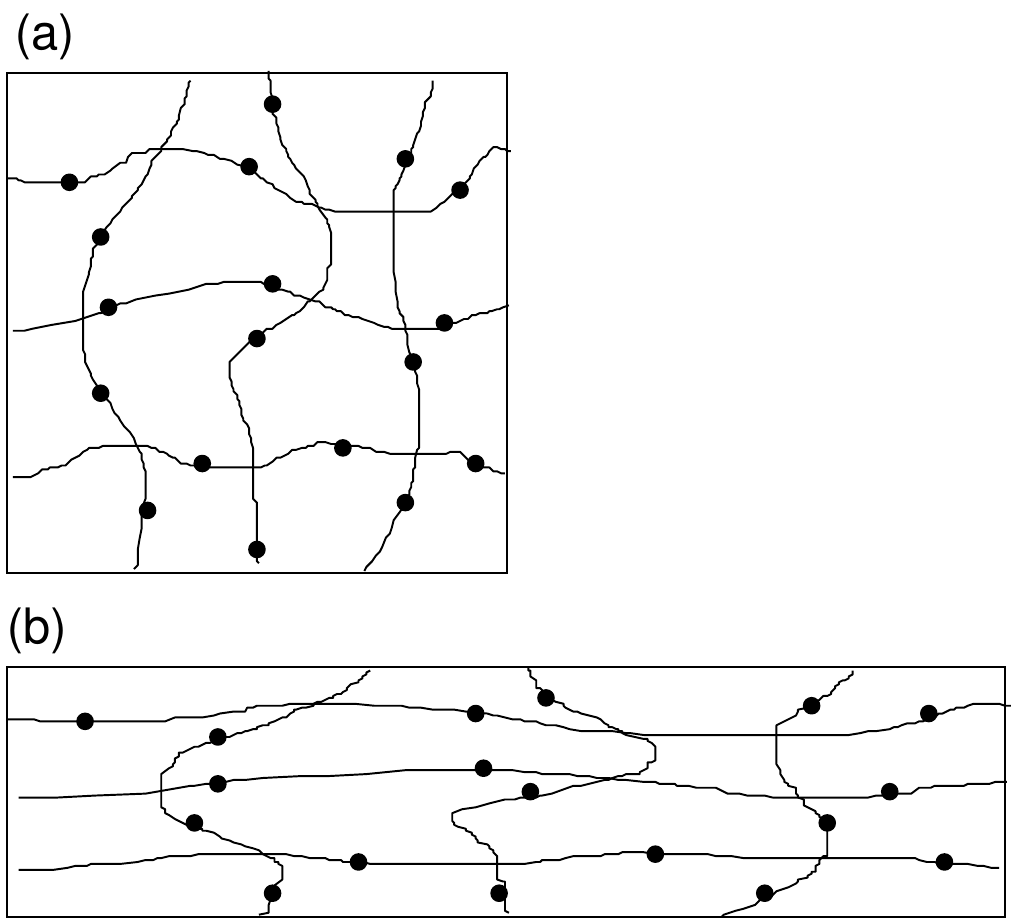}
\caption{\label{path.pdf}
(a) Percolating fluid flow channels (lines) and critical constrictions (black dots) for a $L\times L$ square unit 
system with isotropic roughness.
(b) The percolating fluid flow channels and critical constrictions for a system obtained by stretching by a factor 
of $2$ in the $x$-direction and $1/2$ in the y-direction (Pekeling number $\gamma = 4$). After this mapping, 
the concentration of flow channels is
increased by a factor of 2  in the $x$-direction and reduced by a factor of 1/2 in the $y$-direction. For a square unit
$L\times L$ (not shown) the number of critical constrictions along each percolating flow channel
is reduced by a factor of 1/2 in the $x$-direction and increased by a factor of 2 in the $y$-direction. 
The net result is that the fluid flow conductivity is increased by a factor of 4 in the $x$-direction and reduced by a
factor of 1/4 in the $y$-direction, i.e., $\sigma_x = \gamma \sigma_0$ and $\sigma_y =  \sigma_0/\gamma$, where $\sigma_0$
is the flow conductivity for the system with isotropic roughness in (a).
}
\end{figure}

\begin{figure}
\includegraphics[width=0.45\textwidth,angle=0]{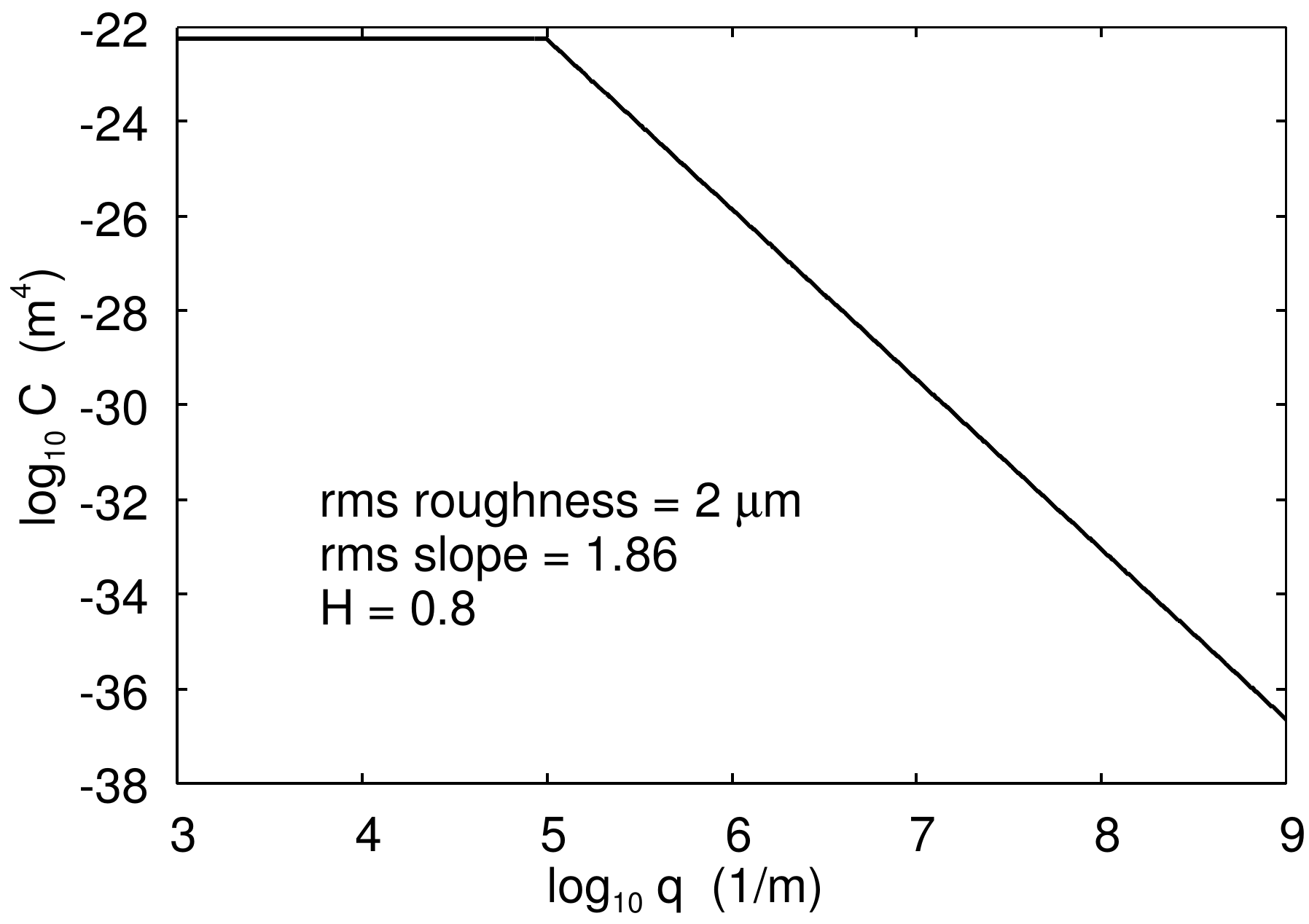}
\caption{\label{1logq.2logC.used.pdf}
Surface roughness power spectrum as a function of the wavenumber (log-log scale).
}
\end{figure}

\begin{figure}
\includegraphics[width=0.45\textwidth,angle=0]{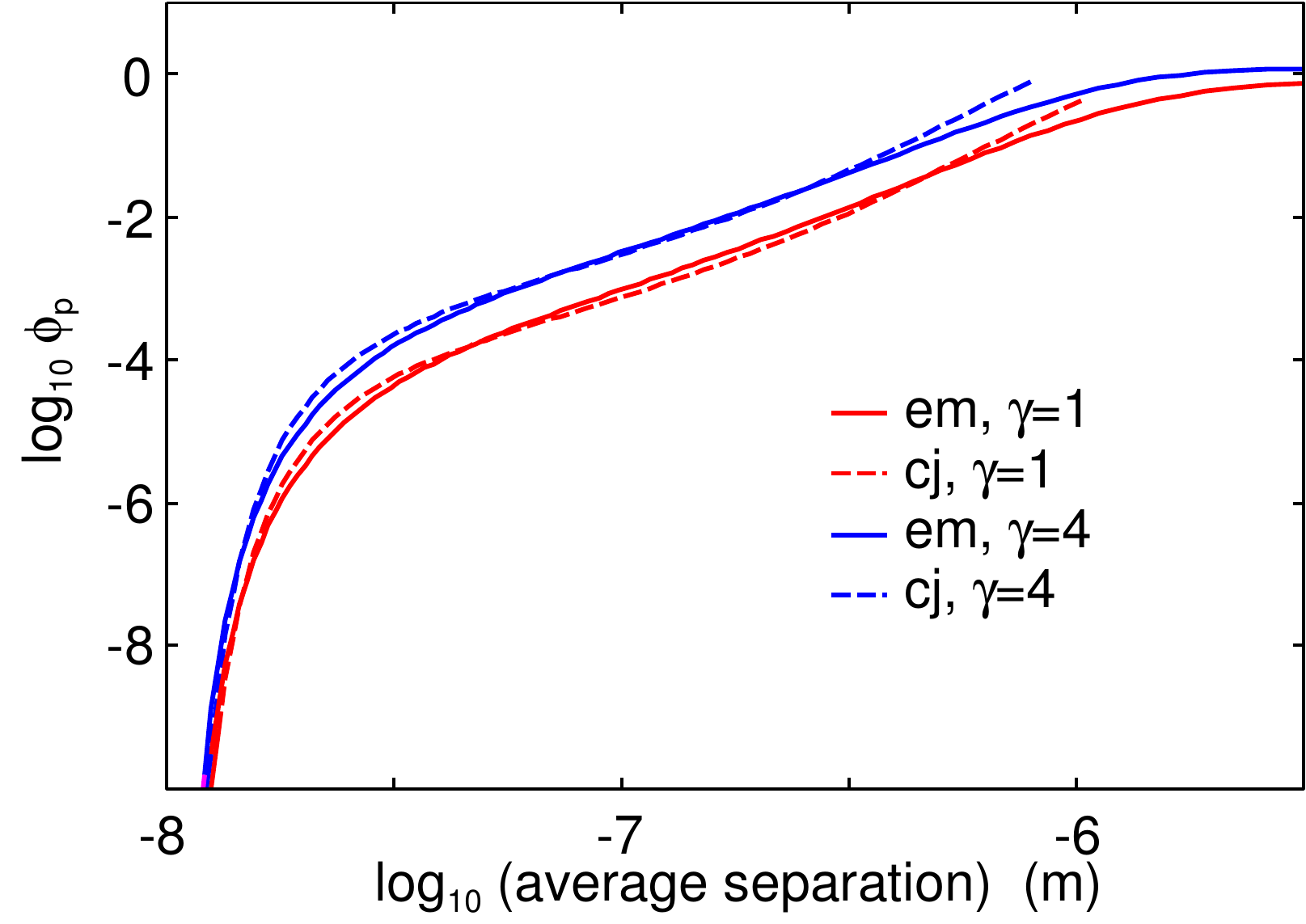}
\caption{\label{1logubar.logphi.pdf}
Fluid pressure flow factor $\phi_{\rm p} =12 \eta \sigma_x/\bar u^3$ 
as a function of the average surface separation $\bar u$ (log-log scale).
In the calculation we have used the surface roughness power spectra shown in Fig. \ref{1logq.2logC.used.pdf}
and the Young's elastic modulus $E=10 \ {\rm MPa}$.
Results are shown for $\gamma=1$ (red curves) and $\gamma=4$ (blue curves) using the effective medium (em) theory (solid lines)
and the critical junction (cj) theory (dashed curves).}
\end{figure}

\vskip 0.3cm
{\bf 6 The critical junction theory of fluid flow}

Consider a rubber seal.
Assume first isotropic roughness and that the nominal contact region between the rubber and the hard counter-surface is a square 
area $L\times L$. We assume that a high-pressure fluid region occur for $x<0$ and a low-pressure region for $x>L$. 
Now, let us study the contact between the two solids as we 
increase the magnification $\zeta$. We define $\zeta=L/\lambda$, where $\lambda$ is the resolution. 
We study how the apparent contact area (projected on the $xy$-plane), $A(\zeta)$, between the two solids depends on the magnification $\zeta$. 
At the lowest magnification we cannot observe any surface roughness, and the contact between the solids appears to be complete i.e., $A(1)=A_0$. 
As we increase the magnification we will observe some interfacial roughness, and the (apparent) contact area will decrease. 
At high enough magnification, say $\zeta=\zeta_{\rm c}$, a percolating path of non-contact area will be observed for the first time. 
We denote the most narrow constriction along this percolation path as the critical constriction. The critical constriction will have the 
lateral size $\lambda_{\rm c} = L/\zeta_{\rm c}$ and the surface separation at this point is denoted by $u_{\rm c}$. 
We can calculate $u_{\rm c}$ using a recently developed contact mechanics theory\cite{YangP}. 
As we continue to increase the magnification we will find more percolating channels between the surfaces, 
but these will have more narrow constrictions than the first channel which appears at $\zeta=\zeta_{\rm c}$, 
and as a first approximation one may neglect the contribution to the leak rate from these channels. 
An accurate estimate of the leak rate is obtained by assuming that all the leakage 
occurs through the critical percolation channel, and that the whole pressure 
drop $\Delta P=P_{\rm a}-P_{\rm b}$ (where $P_{\rm a}$ and $P_{\rm b}$ 
is the pressure to the left and right of the seal) occurs over the critical constriction 
(of width and length $\lambda_{\rm c}\approx L/\zeta_{\rm c}$ and height $u_{\rm c}$). 
We refer to this theory as the critical-junction theory. 
If we approximate the critical constriction as a pore with rectangular cross-section (width and length $\lambda_{\rm c}$ and 
height $u_{\rm c}<< \lambda_{\rm c}$), and if we assume an incompressible Newtonian fluid, the volume flow per unit time through the critical
constriction will be given by (Poiseuille flow) 
$$\dot Q={u_{\rm c}^3 \over 12 \eta}\Delta P\eqno(37)$$ 
In deriving (37) we have assumed laminar flow and that $u_{\rm c} << \lambda_{\rm c}$, which is always satisfied in practice. 
The flow conductivity $\sigma_{\rm eff}$ can be obtained from $\dot Q$ using $\dot Q=J_x L =  \sigma_{\rm eff} (\Delta P/L) L$ giving
$$\sigma_{\rm eff} = {u_{\rm c}^3 \over 12 \eta}\eqno(38)$$

The following qualitative picture underpin the critical constriction model. 
At the critical magnification several fluid conducting channels may appear and each of them 
may have several critical constrictions as indicated in Fig. \ref{path.pdf}(a). 
Now when we perform the mapping indicated in Fig. \ref{percolation.pdf}, where we go from isotropic roughness in
a square area $L\times L$ to the anisotropic roughness in a square area of the 
same size [dashed square in Fig. \ref{percolation.pdf}(c)], 
we increase the number of flow channels in the $x$-direction 
by a factor of $\gamma^{1/2}$, and on each flow channel
we reduce the number of critical junctions by
a factor of $\gamma^{-1/2}$. Hence the the fluid conductivity $\sigma_x = \gamma \sigma_0$.
In a similar way one can show that $\sigma_y = \sigma_0/\gamma$. 
Note that this imply $\sigma_x = \gamma^2 \sigma_y$, which we 
derived above from the effective medium theory close to the contact area percolation threshold.
Note that this agreement with the effective medium theory require that the fluid pressure drop over a
critical constriction is not modified by the stretching--contraction of the system.

To illustrate the accuracy of the critical junction approach, 
In Fig. \ref{1logubar.logphi.pdf} I show the fluid pressure flow factor $\phi_{\rm p} =12 \eta \sigma_x/\bar u^3$ 
as a function of the average surface separation $\bar u$ (log-log scale).
In the calculation we have used the surface roughness power spectra shown in Fig. \ref{1logq.2logC.used.pdf}
and the Young's elastic modulus $E=10 \ {\rm MPa}$.
Results are shown for $\gamma=1$ (red curves) and $\gamma=4$ (blue curves) using the effective medium theory (solid lines)
and the critical junction theory (dashed curves). As expected, the critical junction theory is accurate when the average surface separation
is small enough but is inaccurate for very small contact pressures where the average surface separation is large; this is expected as for
large average surface separation a nearly uniformly thick fluid film separate the surfaces and the fluid pressure drop will
not occur over a small number of narrow constrictions, but will occur in nearly uniformly over the whole nominal contact area.
However, this limiting case is not of interest in sealing applications.


\vskip 0.3cm
{\bf 7 Discussion}

In Ref. \cite{JPC} we used molecular dynamic simulations 
to study the percolation of the contact area
with increasing pressure for Tripp numbers $0.5 < \gamma < 2$.
We found that the results could be reasonably well fit with the formula
$A/A_0 = \gamma /(1+\gamma)$. However, the Bruggeman effective medium theory
predict $A/A_0 = 0.5$ independent of $\gamma$, and it is clear from very simple
arguments (see Sec. 2) that for an infinite system the percolation threshold does
not depend on the asymmetry (or stretching) parameter $\gamma$. Hence, the fact that
the numerical simulations showed a dependency of $A/A_0$ on $\gamma$ 
must be a finite-size effect.

Recently, Yang et al have performed a numerical study of the effect of 
surface roughness anisotropy on the percolation threshold of sealing surfaces\cite{A1}.
For surfaces with isotropic roughness they found $A/A_0 \approx 0.48$, i.e., close the the 
effective medium theory prediction, and larger than the value $0.42$ found by Dapp et al\cite{Dapp}.

As  $\gamma$ increased from $0.5$ to $1.66$, Yang et al found that $A/A_0$ 
increased from $0.43$ to $0.53$, which is a weaker $\gamma$-dependency then given by $A/A_0=\gamma/(1+\gamma)$,
which predict that $A/A_0$ increases from $0.33$ to $0.63$. This result is expected since
in the limit of an infinite system the percolation threshold is independent of $\gamma$,
and for finite systems the $\gamma$-dependency must depend on the system size.

The good agreement found between the effective medium theory and the critical junction theory indicate that the basic picture
behind the critical junction theory is accurate. The critical junction theory is based on the observation that
when increasing the magnification, at high enough magnification, say $\zeta=\zeta_{\rm c}$, 
a percolating path of non-contact area will be observed for the first time. 
As we continue to increase the magnification we find more percolating channels between the surfaces, 
but these will have more narrow constrictions than the first channel which appears at $\zeta=\zeta_{\rm c}$, 
and as a first approximation one may neglect the contribution to the leak rate from these channels. 
This imply that the roughness observed when the magnification is 
increased beyond $\zeta=\zeta_{\rm c}$ has a negligible influence on the 
leakage of a seal. I a recent comment, Papangelo et al\cite{A2} 
state that the leakage rate depends on the short distance cut-off length $\lambda_1$ (observed at the
highest magnification $\zeta_1$), which could be an atomic distance,
but this is in general not the case unless then nominal 
pressure is so high as to move the critical constriction
to the shortest length scale, which is nearly never the case in practical applications.

\vskip 0.3cm
{\bf 8 Summary and conclusion}

I have studied the influence of anisotropic roughness on the fluid flow at the interface between 
two elastic solids. I have shown that for randomly rough surfaces with anisotropic roughness, the contact area
percolate at the same relative contact area as for isotropic roughness, and that
the Bruggeman effective medium theory and the critical junction theory gives nearly the
same results for the fluid flow conductivity (and the fluid pressure flow factor).
This shows (qualitatively) that, unless the nominal contact pressure
is so high at to result in nearly complete contact, 
the surface roughness observed at
high magnification is irrelevant for the fluid flow during squeeze-out, or for the
leakage of stationary seals.

\vskip 0.3cm
{\bf Acknowledgments:}
I thank M. Scaraggi for useful comments on the manuscript!

\vskip 0.3cm
{\bf Appendix A: Two integrals}

In Sec. 3 and 4 appeared two important integrals:

$$I_1 = {1\over 2 \pi} \int_0^{2 \pi} d \phi \ {{\rm cos}^2\phi \over {\rm cos}^2\phi + \gamma^2 {\rm sin}^2\phi}= 
{1 \over 1+\gamma}\eqno(A1)$$

$$I_2 = {1\over 2 \pi} \int_0^{2 \pi} d \phi \ {{\rm sin}^2\phi \over {\rm cos}^2\phi + \gamma^2 {\rm sin}^2\phi}= 
{1 \over \gamma (1+\gamma)}\eqno(A2)$$

Note that $I_1+\gamma^2 I_2 = 1$. The results above can the proved by integration in the complex plane: Introducing $z=e^{i\phi}$ and writing
${\rm cos}\phi = (z+1/z)/2$ and ${\rm sin}\phi = (z-1/z)/(2i)$ and performing the integration around the circle $|z|=1$ result in the
equations (A1) and (A2).

\vskip 0.3cm
{\bf Appendix B: Flow current in elliptic insertion}

The 2D fluid flow problem is mathematical similar to the 
electrostatic polarization of a dielectric material. Thus 
the fluid flow current and the electric current both satisfies
$\nabla \cdot {\bf J}=0$ (conservation of fluid volume and electric charge,
respectively). The fluid current is related to the pressure gradient via ${\bf J}=-\sigma \nabla p$,
where $\sigma$ is the flow conductivity,
and the electric current is related to the electric potential via ${\bf J}=-\sigma \nabla \phi$,
where $\sigma$ is the electric conductivity. Hence, results obtained in electrostatics for
polarizable media can be used also for the fluid flow problem. In particular, from electrostatics it
is known that if an elliptic region
with constant dielectric properties is embedded in an infinite dielectric material with other
dielectric properties, then the electric field (and hence the polarization) 
in the elliptic region will be constant, assuming that the
electric field is constant far away from the elliptic region. The corresponding 
result for the fluid flow problem was used in Sec. 4.

Note that when $\sigma$ is constant the equation $\nabla \cdot {\bf J}=0$ gives $\nabla^2 p =0$ (for fluid flow)
and $\nabla^2 \phi =0$ (for electrostatics).
The results for the electric polarization problem for an elliptic insertion can be derived by solving 
the Laplace equation $\nabla^2 \phi =0$ using elliptic coordinates\cite{Mors} or by 
complex mapping methods\cite{complex}.

%

\end{document}